\documentclass[amsmath,amssymb,floatfix,groupedaddress,twocolumn,prl]{revtex4}
\usepackage{graphicx}
\usepackage{color}

\newcommand{\msub}[1]{_{\mathrm{#1}}}
\newcommand{\msup}[1]{^{\mathrm{#1}}}

\begin{document}

\title{Wavelength- and material-dependent absorption in GaAs and AlGaAs microcavities}

\author{C. P. Michael}
\email{cmichael@caltech.edu}
\affiliation{Department of Applied Physics, California Institute of Technology, Pasadena, CA 91125}
\author{K. Srinivasan}
\altaffiliation{Also at:  Center for the Physics of Information, California Institute of Technology, Pasadena, CA 91125}
\affiliation{Department of Applied Physics, California Institute of Technology, Pasadena, CA 91125}
\author{T. J. Johnson}
\affiliation{Department of Applied Physics, California Institute of Technology, Pasadena, CA 91125}
\author{O. Painter}
\affiliation{Department of Applied Physics, California Institute of Technology, Pasadena, CA 91125}
\author{K. H. Lee}
\altaffiliation{Also at:  Department of Physics, University of Oxford, Parks Road, Oxford, OX1 3PU, United Kingdom}
\affiliation{California Nanosystems Institute, University of California, Santa Barbara, CA 93106}
\author{K. Hennessy}
\affiliation{California Nanosystems Institute, University of California, Santa Barbara, CA 93106}
\author{H. Kim}
\affiliation{California Nanosystems Institute, University of California, Santa Barbara, CA 93106}
\author{E. Hu}
\affiliation{California Nanosystems Institute, University of California, Santa Barbara, CA 93106}

\date{\today}

\begin{abstract} 
The quality factors of modes in nearly identical GaAs and Al$_{0.18}$Ga$_{0.82}$As microdisks are tracked over three wavelength ranges centered at 980\,nm, 1460\,nm, and 1600\,nm, with quality factors measured as high as  6.62$\times$10$^5$ in the 1600-nm band.  After accounting for surface scattering, the remaining loss is due to sub-bandgap absorption in the bulk and on the surfaces.  We observe the absorption is, on average, 80\,percent greater in AlGaAs than in GaAs and in both materials is 540\,percent higher at 980\,nm than at 1600\,nm.
\end{abstract}

\maketitle

In recent semiconductor cavity QED experiments involving self-assembled III-V quantum dots (QDs), Rabi splitting of the spontaneous emission line from individual QD excitonic states has been measured for the first time~\cite{nat432-197,nat432-200}.  Potential application of these devices to quantum networks~\cite{prl78-3221} and cryptography~\cite{nat420-762} over long-haul silica fibers has sparked interest in developing QD-cavity systems with efficient light extraction operating in the telecommunication bands at 1300\,nm and 1550\,nm~\cite{jjap44-L620}.  The demonstration of vacuum-Rabi splitting in this system, a result of coupling a QD to localized optical modes of a surrounding microresonator, has been greatly aided by prior improvements to the design and fabrication of semiconductor microcavities~\cite{nat425-944,apl86-111105,oe14-3472}.  At the shorter wavelengths involved in these Rabi splitting experiments (740--1200\,nm), the optical quality factors ($Q$) of the host AlGaAs microcavities were limited to $Q \approx $ 2$\times$10$^4$---corresponding to a loss rate comparable with the coherent QD-cavity coupling rate.  Further reduction of optical loss would increase the relative coherence of the QD-cavity system and would allow greater coupling efficiency to the cavity mode. 

In previous measurements of wavelength-scale AlGaAs microdisk resonators, we have demonstrated $Q$-factors up to 3.6$\times$10$^5$ in the $1400$-nm band~\cite{apl86-151106} and attributed the improved performance to an optimized resist-reflow and dry-etching technique, which produces very smooth sidewalls~\cite{oe13-1515}.  Subsequently, we have also measured Al$_{0.3}$Ga$_{0.7}$As microdisks with similar quality factors between 1200\,nm and 1500\,nm; however, these disks exhibit a significant unexpected decrease in $Q$ at shorter wavelengths ($\lambda\msub{o} \approx 852$\,nm)~\cite{qels2005-Kartik}.  In related work on silicon microdisks, methods have been developed to specifically measure and characterize losses due to material absorption and surface scattering~\cite{apl85-3693,oe13-1515,apl88-131114}.  In this Letter we study the properties of GaAs and Al$_{0.18}$Ga$_{0.82}$As microdisks across three wavelength bands centered at 980\,nm, 1460\,nm, and 1600\,nm.  After estimating and removing the surface-scattering contribution to the cavities losses, we find the remaining absorption, composed of losses in the bulk and on the surfaces, depends significantly on both wavelength and material composition.  


Within the microdisk resonators studied here, optical loss can be separated into three main components: intrinsic radiation of the whispering-gallery modes (WGMs), scattering from roughness at the air-dielectric interface due to fabrication imperfections, and absorption at the surface or in the bulk of the semiconductor material.  The measured total intrinsic $Q\msub{i}$ is given by
\begin{equation}
1/Q\msub{i} = 1/Q\msub{rad} + 1/Q\msub{ss} + 1/Q\msub{a},
\label{totalQ}
\end{equation}
where $Q\msub{rad}$, $Q\msub{ss}$, and $Q\msub{a}$ describe cavity losses to radiation, surface scattering, and absorption, respectively.  As in the devices studied here for disk diameters (thicknesses) $\gg${}$\lambda\msub{o}/n\msub{d}$ ($>${}$\lambda\msub{o}/2n\msub{d}$) where $\lambda\msub{o}$ is the free-space wavelength and $n\msub{d}$ is the refractive index of the disk material, microdisk cavities support a large number of modes with very low radiation loss.  For all the microdisk modes studied here, the calculated $Q\msub{rad}$ is $\gtrsim$10$^6$ and typically is $>$10$^8$. 

The samples were fabricated from high-quality heterostructures grown by molecular beam epitaxy (MBE) on a GaAs substrate.  Two different samples were grown: a ``GaAs'' sample containing a 247-nm GaAs disk layer, and an ``AlGaAs'' sample with a 237-nm Al$_{0.18}$Ga$_{0.82}$As disk layer.  In both samples the disk layer was grown nominally undoped (background doping levels $n_p \lesssim 10^{15}$\,cm$^{-3}$) and was deposited on a 1.6-$\mu$m Al$_{0.7}$Ga$_{0.3}$As sacrificial layer.  Microdisks with a  radius of $\sim$3.4\,$\mu$m were defined by electron-beam lithography and etched in a 55~percent (by volume) HBr solution containing 3.6\,g of K$_2$Cr$_2$O$_7$ per litre~\cite{apl75-1908}.  The disks were partially undercut by etching the sacrificial layer in 8\,percent HF acid for 45\,s, prior to e-beam resist removal.
\begin{figure}
\includegraphics[width = \columnwidth]{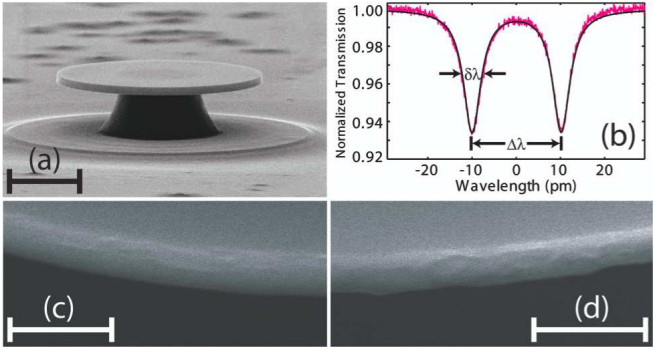}
\caption{(a) SEM image of a typical GaAs microdisk.  (b) Sample scan of a TE$_{p=4}$ doublet in a GaAs disk at $\lambda\msub{o} = 1582.796$\,nm.  Fitting the data gives the linewidth $\delta\lambda = 4.57$\,pm and the splitting $\Delta\lambda = 20.20$\,pm.  Although not visible in (a), the edge roughness is different for the (c) GaAs and (d) AlGaAs samples.  Scale bars are 2.5\,$\mu$m in (a) and 250\,nm in (c) and (d).}
\label{SEM_samples}
\end{figure}

Passive measurements of the cavity $Q\msub{i}$ were performed using an evanescent coupling technique employing a dimpled fiber-taper waveguide~\cite{apl85-3693,prb70-081306R,CM_dimple_taper}.  The dimpled taper was mounted to a three-axis 50-nm-encoded stage and positioned in the near field of the resonator.  Using the taper position to vary the coupling, the WGMs of the microdisks were excited using three swept tunable laser sources (linewidth $< 5$\,MHz, covering 963--993\,nm, 1423--1496\,nm, and 1565--1625\,nm).  By weakly loading the cavity, the resonance linewidth $\delta\lambda$ [see Fig.~\ref{SEM_samples}(b)] is a good measure for the intrinsic quality factor ($Q\msub{i} = \lambda\msub{o}/\delta\lambda$).
\begin{figure}
\includegraphics{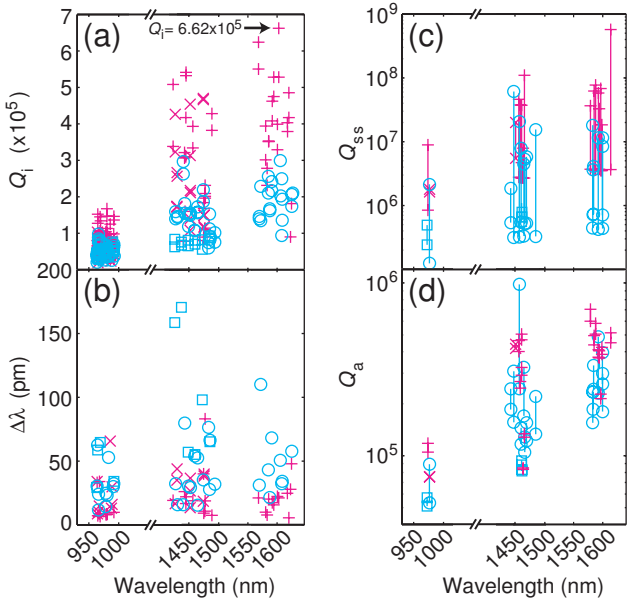}
\caption{Measured  (a) $Q\msub{i}$ and (b) $\Delta\lambda$ for the (\textcolor{magenta}{$+$},\textcolor{magenta}{$\times$}) GaAs TE- and TM-polarized microdisk modes and (\textcolor{cyan}{{\footnotesize $\bigcirc$}},\textcolor{cyan}{$\square$}) Al$_{0.18}$Ga$_{0.82}$As TE and TM modes, respectively.  In (c) and (d), connected points represent calculated bounds on  (c) $Q\msub{ss}$ and (d) $Q\msub{a}$ for each family of modes.  The data were compiled from two disks of each material.}
\label{Qs_graph}
\end{figure}
High-resolution linewidth scans were calibrated to $\pm0.01$\,pm accuracy using a fiber-based Mach-Zehnder interferometer.  A set of polarization-controlling paddle wheels was used to selectively couple to TE-like or TM-like microdisk modes.  Each family of modes (same radial order $p$) was identified by comparing the coupling behavior and free spectral range (FSR) to finite-element method (FEM) models~\cite{apl88-131114}.

The measured $Q\msub{i}$s for all observed modes are summarized in Fig.~\ref{Qs_graph}(a).  Modes with $Q\msub{i}$s dominated by radiation loss, {\frenchspacing i.e.} the measured $Q\msub{i}$ is near $Q\msub{rad}$ calculated using FEM simulations, are omitted.  In the 1600-nm band, the high-$Q$ TE modes are $p =$ 1--4 in GaAs and $p=$ 1--3 in AlGaAs; all TM modes are radiation limited in this band.  Near 1460\,nm, the TE$_{p=1-4}$ and TM$_{p=1}$ modes in both materials are detectable and not radiation limited.  In the 980-nm range, identifying modes becomes more difficult:  at this wavelength families through TE$_{p=8}$ and TM$_{p=7}$ are not radiation limited, and significant spectral overlap between the modes causes Fano-like resonance features~\cite{pr124-1866}.  In addition, we are unable to couple to the lowest order modes of both polarizations ($p\approx$ 1--3) because they are poorly phase matched to the fiber taper.  

Despite efforts to produce perfectly smooth side walls, Figs.~1(c,d) indicate that significant surface roughness is still present.  Surface roughness on microdisk resonators backscatters light between the degenerate WGMs, which breaks their degeneracy and results in the measured mode-splitting ($\Delta\lambda$) shown in Fig.~\ref{SEM_samples}(b).   Following the theory developed in Refs.~\cite{oe13-1515} and~\cite{ol21-1390},
$\Delta\lambda$ and $Q\msub{ss}$ are both dependent on the characteristic volume of the scatterer ($V\msub{s}$):
\begin{eqnarray}
\Delta\lambda & = & \frac{\pi^{3/4}}{\sqrt{2}}\lambda\msub{o}V\msub{s}(n\msub{d}^2-n\msub{o}^2)\sum_{\hat{\eta}}\overline{u}\msub{s}(\hat{\eta}) \\
Q\msub{ss} & = & \frac{\lambda\msub{o}^3}{\pi^{7/2}n\msub{o}(n\msub{d}^2-n\msub{o}^2)^2V\msub{s}^2\sum_{\hat{\eta}}\overline{u}\msub{s}(\hat{\eta})G(\hat{\eta})}
\end{eqnarray}
where $n\msub{d}$ and $n\msub{o}$ are the indices of refraction of the disk and surrounding medium, respectively, $\overline{u}\msub{s}(\hat{\eta})$ is the spatially-averaged $\hat{\eta}$-polarized normalized electric field energy at the disk edge, and $G(\hat{\eta}) = \{2/3,2,4/3\}$ is a geometrical factor weighting the radiation contribution from the $\hat{\eta} = \{\hat{r},\hat{\phi},\hat{z}\}$ polarizations.   
The mode field profiles are calculated by the FEM.  For FEM models in the 980-nm span, we treat all measured TE (TM) modes as TE$_{p=7}$ (TM$_{p=6}$) because the appropriate field parameters do not vary significantly between radial orders.  

We employ two separate measurements to find rough bounds on $Q\msub{ss}$.  First, we use the average doublet splitting for each family to find the average ${\langle}V\msub{s}{\rangle}_{p}$ sampled by each mode and then calculate the $Q\msub{ss}$ associated with each family.
$$\textrm{Splitting Method:\ \ }  \Delta\lambda \Rightarrow {\langle}V\msub{s}{\rangle}_{p} \Rightarrow Q\msub{ss}$$  
Second, we statistically analyze the roughness of the disk edges in high resolution SEM images~\cite{apl85-3693}.  Fitting the autocorrelation of the roughness to a Gaussian, the roughness amplitude ($\sigma\msub{r}$) and correlation length ($L\msub{c}$) give the ``statistical'' scatterer volume ($\overline{V}\msub{s} = \sigma\msub{r}t\sqrt{RL\msub{c}}$ where $t$ and $R$ are the disk's thickness and radius) for each disk, which is used to estimate $Q\msub{ss}$.
$$\textrm{Statistical Method:\ \ }  \sigma\msub{r},L\msub{r} \Rightarrow \overline{V}\msub{s} \Rightarrow Q\msub{ss}$$
The average $\{\sigma\msub{r},L\msub{c}\}$ for the GaAs and AlGaAs disks are $\{0.6,38.7\}$\,nm and $\{1.8,29.4\}$\,nm, respectively.  Because each mode will not sample all of the disk's physical irregularities, the roughness estimated by the statistical analysis is slightly greater than the roughness calculated from the doublet splittings.   Hence, the  doublet splitting places an upper bound and more accurate value for $Q\msub{ss}$ [Fig.~\ref{Qs_graph}(c)]. The statistical analysis gives a lower bound, although neither bound is strict in the theoretical sense.

Through Eq.~(\ref{totalQ}), $Q\msub{ss}$ and $Q\msub{rad}$ are removed from the measured $Q\msub{i}$ to obtain limits on $Q\msub{a}$ [Fig.~\ref{Qs_graph}(d)].  
To relate cavity losses to material properties, the material absorption rate ($\gamma_{\mathrm{a},p}$) for the $p\msup{th}$~mode is given by $\gamma_{\mathrm{a},p} = 2\pi{}c/\lambda\msub{o}Q\msub{a}$~\footnote{The commonly used absorption coefficient ($\alpha$) depends on both the material absorption rate and the group velocity of the cavity mode ($\alpha = \gamma\msub{a}/v\msub{g}$).}.  
We weight each measured doublet equally and average over all families in a band to determine an average $\overline\gamma\msub{a}$. Table~\ref{gamma_table} compiles the average absorption rates for both GaAs and Al$_{0.18}$Ga$_{0.82}$As across the three wavelength ranges.  
\begin{table}
\caption{Summary of material absorption rates.}
\label{gamma_table}
\begin{ruledtabular}
\begin{tabular}{l c c}
Method:     		& 	Splitting			& 	Statistical 			\\ 
\hline
Sample     	& \multicolumn{2}{c}{$(\overline\gamma\msub{a}/2\pi)\pm1\sigma$ (GHz)}	\\
\hline
$\begin{array}{l} \mbox{GaAs} \\ \quad @\ 980\,\mbox{nm}\,\  \end{array} $	
	$\Bigg\{ \begin{array}{c} \mbox{all modes} \\ \mbox{TE} \\ \mbox{TM} \end{array} $
	&	$ \begin{array}{c}  3.47\pm0.59 \\ 2.94\pm1.24 \\ 4.08\pm0.91 \end{array} $
	&	$ \begin{array}{c}  3.29\pm0.75 \\ 2.61\pm1.28 \\ 4.06\pm0.89 \end{array} $		\\
$\begin{array}{l} \mbox{AlGaAs} \\ \quad @\ 980\,\mbox{nm}\,\  \end{array} $ 
	$\Bigg\{ \begin{array}{c} \mbox{all modes} \\ \mbox{TE} \\ \mbox{TM} \end{array} $		
	&	$ \begin{array}{c}  5.84\pm0.13 \\ 5.77\pm1.61 \\ 6.03\pm1.56 \end{array} $
	&	$ \begin{array}{c}  4.00\pm0.91 \\ 3.44\pm2.31 \\ 5.39\pm1.73 \end{array} $		\\
$\begin{array}{l} \mbox{GaAs} \\ \quad @\ 1460\,\mbox{nm} \end{array} $
	$\Bigg\{ \begin{array}{c} \mbox{all modes}\\ \mbox{TE}_{p=1} \\ \mbox{TM}_{p=1}  \end{array} $		
	&	$\begin{array}{c}  0.942\pm0.696 \\ 0.514\pm0.085 \\ 0.495\pm0.089 \end{array} $
	&	$\begin{array}{c}  0.888\pm0.692 \\ 0.444\pm0.108 \\ 0.467\pm0.095 \end{array} $	\\
$\begin{array}{l} \mbox{AlGaAs} \\ \quad @\ 1460\,\mbox{nm} \end{array} $
	$\Bigg\{ \begin{array}{c} \mbox{all modes} \\ \mbox{TE}_{p=1} \\ \mbox{TM}_{p=1} \end{array} $		
	&	$\begin{array}{c}  1.73\pm0.50 \\ 1.32\pm0.22    \\ 2.30\pm0.49 \end{array} $
	&	$\begin{array}{c}  1.43\pm0.76 \\ 0.882\pm0.380 \\ 2.39\pm0.40 \end{array} $	\\
GaAs @ 1600\,nm\ \ \,-- TE	&		$0.507\pm0.186$		&	$0.460\pm0.185$\\
AlGaAs\,@\,1600\,nm\,-- TE &		$0.968\pm0.179$		&	$0.629\pm0.173$	\\
\end{tabular}
\end{ruledtabular}
\end{table}
The average rates are 540\,percent larger at 980\,nm than at 1600\,nm and 80\,percent greater in AlGaAs than in GaAs.  

The measured absorption may be due to a number of sources.  Although nonlinear-absorption-induced optical bistability was measured for internal cavity energies as low as 106\,aJ, the losses reported in Table \ref{gamma_table} were all taken at input powers well below the nonlinear absorption threshold.  Free carrier absorption can also be neglected given the nominally undoped material and relatively short wavelengths studied here~\cite{pr114-59}.  The Urbach tail makes a small contribution in the 980-nm band ($\leq$15\,percent) and is negligible otherwise~\cite{pr114-59}.  This leaves deep electron (hole) traps as the major source contributing to bulk material absorption in the measured microdisks.  Similar wavelength dependent absorption has been observed in photocurrent measurements of MBE-grown AlGaAs waveguides~\cite{ieee-jqe33-933} and attributed to sub-bandgap trap levels associated with vacancy complexes and oxygen incorporation during growth~\cite{jap61-5062}.  Given the high surface-volume ratio of the microdisks, another possible source of loss is surface-state absorption.  The sensitivity to absorption from surface-states can be quantified by the $p\msup{th}$ mode's energy overlap with the disk's surface, $\Gamma'\msub{p}$; TM modes are more surface-sensitive than TE modes~\cite{apl88-131114} whereas both polarizations are almost equally sensitive to the bulk.  The calculated surface overlap ratio is $\Gamma'\msub{TM}/\Gamma'\msub{TE} \approx 2.65$ for $p=1$ modes in the 1460-nm band, where all surfaces of the disk (top, bottom, and etched edge) are treated equally.  For these modes the measured absorption ratio is $\gamma\msub{a,TM}/\gamma\msub{a,TE} = 1.74\pm0.47$ ($0.96\pm0.23$) in the AlGaAs (GaAs) microdisks, which indicates the presence of significant surface-state absorption in the AlGaAs resonators and dominant bulk absorption in the GaAs disks.  In the 980-nm band, the data are consistent with bulk absorption [$\gamma\msub{a,TM}/\gamma\msub{a,TE}=1.05\pm0.40$ ($1.39\pm0.66$) for the AlGaAs (GaAs) devices] although the results are less conclusive due to the larger scatter in the data.

In summary, after accounting for radiation and surface scattering losses, we measure greater sub-bandgap absorption in Al$\msub{0.18}$Ga$\msub{0.82}$As microdisks than in similar GaAs resonators, and the absorption in both materials decreases towards longer wavelengths.  From the polarization dependence of the measured optical loss, we infer that both surface states and bulk states contribute to the residual absorption in these structures.  Our results imply that reductions in the optical loss of AlGaAs-based microphotonics, especially at the shorter wavelengths $< 1$\,$\mu$m and in high Al content alloys, will require further study and reduction of deep level traps, and that surface passivation techniques~\cite{apl64-1911} will also likely be important.

Two authors (C.P.M. and K.S., respectively) would like to thank the Moore Foundation and the Hertz Foundation for their graduate fellowship support.  K. H. L.  thanks the Wingate Foundation.  H. K. has been supported by NSF grant No.~0304678.

\end{document}